\documentclass[aps,prl,twocolumn,superscriptaddress,showpacs,10pt]{revtex4-1}
\usepackage{natbib}
\usepackage{mathbbol,amsmath,amsfonts,amssymb,bm,color}
\usepackage{graphicx,psfrag,array,braket}
\usepackage{fixltx2e,float}

\providecommand{\ignore}[1]{}
\providecommand{\aucmnt}[1]{#1}
\def\be{\begin{equation}}
\def\ee{\end{equation}}
\renewcommand{\aucmnt}[1]{}

\newcommand{\Comment}[1]{}

\newcommand{\Eq}[1]{Eq.~(\ref{#1})}

\begin{document}

\title{Entangled Pauli Principles: the DNA of Quantum Hall Fluids}
\author{Sumanta Bandyopadhyay}

\affiliation{Department of Physics, Washington University, St. Louis, MO 63130, USA}
\author{Li Chen}
\affiliation{National High Magnetic Field Laboratory and Department of Physics, Florida State University, Tallahassee, FL 32306, USA}

\author{Mostafa Tanhayi Ahari}
\affiliation{Department of Physics, Indiana University, Bloomington, IN 47405-7105,USA}
\author{Gerardo Ortiz}

\affiliation{Department of Physics, Indiana University, Bloomington, IN 47405-7105,USA}
\affiliation{Department of Physics, Washington University, St. Louis, MO 63130, USA}
\author{Zohar Nussinov}
\author{Alexander Seidel}
\affiliation{Department of Physics, Washington University, St. Louis, MO 63130, USA}
\date{\today}
\begin{abstract}
A formalism is developed for the rigorous study of  solvable fractional quantum Hall parent Hamiltonians with Landau level mixing. The idea of organization through ``generalized Pauli principles''
is expanded to allow for root level entanglement, giving rise to ``entangled Pauli principles''. Through the latter, aspects of the effective field theory description become ingrained in exact microscopic solutions for a great wealth of phases for which no similar single Landau level description is known.
We discuss in detail braiding statistic, edge theory, and rigorous zero mode counting for the Jain-221 state as derived from a microscopic Hamiltonian. The relevant root-level entanglement is found to feature an AKLT-type MPS structure associated with  an emergent SU(2) symmetry.
\end{abstract}
\pacs{}
\maketitle

\textbf{Introduction.}
The fractional quantum Hall (FQH) regime plays host to an astonishing wealth of interacting topological phases.
A rich theoretical framework describing such phases has historically nucleated around a construction principle for holomorphic lowest Landau level (LL) wave functions \cite{laughlin} and fruitful generalizations to the non-holomorphic, higher LL situation, with optional subsequent lowest LL projection \cite{jain}. This variational principle has proven invaluable in driving the development of field-theoretic descriptions of both the bulk and the edge physics and their intimate relation \cite{MR,wenCS}. One may take the point of view that a complete many-body theory of any correlated phase of matter
requires, in addition to the aforementioned ingredients, a microscopic Hamiltonian granting analytic access to its low energy sector, reproducing key aspects of the field theoretic description of such a phase.
Such ``parent Hamiltonians''  do exist for many \cite{haldane_hierarchy, TK, halperin, gaffnian, Simon} FQH liquids but seem to be lacking for even more. Notably, to our knowledge, they seem to be lacking for most Jain states, which are regarded fundamental both theoretically and experimentally.

In this letter, we argue that the lack of microscopic Hamiltonians stabilizing representative variational wave functions for many
interesting FQH phases is chiefly due to the complexities involved in dealing with non-holomorphic variational states.
This includes unprojected Jain states \cite{jain} as well as more general ``parton'' constructions \cite{jain2, wenparton}.
In these cases, lowest LL projection leads to sufficiently intractable wave functions to preclude the construction of parent
Hamiltonians. Moreover, the unprojected, higher LL variational states are still lacking many of the ``analytic clustering'' properties
that were instrumental in the construction of parent Hamiltonians for many lowest LL states \cite{haldane_hierarchy, TK, halperin}.
For these reasons, even in those cases where parent Hamiltonians have been proposed for higher LL states, rigorous analytic results are usually lacking.
This is in particular true for zero mode counting, from which the case for incompressibility at a certain special filling factor is usually made. Here we will develop principles to study
the zero mode properties of frustration free multiple-LL parent Hamiltonians on the same footing as for similar single-LL
Hamiltonians. Notably, our framework is second-quantized and  {\em de}-emphasizes analytic clustering properties \cite{ortiz}, which are arguably
less useful in the higher LL situation, as we will demonstrate explicitly below. It is worth noting that this lack of emphasis on analytic properties has also been advocated recently by Haldane for somewhat different reasons \cite{haldane11}, and is attractive from the point of view of a description purely in terms of guiding centers (see also \cite{Murthy,Chen14, Mazaheri15, CHLee, kang2017neutral}). Our approach naturally connects with the topical problem of understanding frustration free lattice Hamiltonians and their matrix-product ground states (MPS), with the important additional feature that it deals with non-local such Hamiltonians, and, in principle, MPS of infinite bond dimension \cite{dubail2012edge,MPS_Hall_literature, MPS_Hall_literature1}.

The heart of our framework consists in further elaboration on the concept of a ``generalized Pauli principle'' (GPP), various guises of which have recently played an important role  in discussing the structure of single LL wave functions \cite{ bernevigPRL08, regnault,ThomalePRB84:45127,seidel05,seidel2006abelian, SeidelYang08,bergholtz2006one,bergholtz2006pfaffian,wenwang1,wenwang2}.
Our extension of these ideas not only provides a foundation based on Hamiltonian principles but also generalizes to multiple LLs. The latter will naturally lead to what we coin ``entangled Pauli principles'' (EPPs), which, in addition to the now familiar { rules for GPPs, permit MPS-like entanglement at ``root level'' encoding the quantum fluid's DNA}. We argue this generalization to be key in the endeavor to give a microscopic Hamiltonian description of this type to possibly all FQH phases.
We demonstrate our approach in detail for the parent Hamiltonian of the Jain 221-state \cite{jain221}.
By rigorously establishing the zero mode structure of this Hamiltonian, we make direct contact both with bulk topological and edge conformal properties. As a byproduct, this establishes a case where a simple two-body interaction stabilizes a non-Abelian FQH state, in contrast to better known higher-body, single LL cases \cite{greiter, readrezayi}.

\textbf{Parent Hamiltonian.}
Consider the $n$-Landau level projected ``Trugman-Kivelson'' interaction for fermions,
\begin{equation}\label{eq1}
H_{\sf TK}=\sum_{i<j} P_n\;  \partial_{z_i}\partial_{\bar{z}_i}\delta(z_i-z_j)\delta(\bar{z}_i-\bar{z}_j)\; P_n\,,
\end{equation}
where $z_i=x_i+iy_i$ is the complex coordinate of the $i$th particle, and $\bar{z}_i$ its complex conjugate.
For general projection $P_n$ onto the subspace spanned by the lowest $n$ LLs, this interaction is positive (semi-)definite.
If the $n$-Landau levels are  energetically quenched, as one may choose to assume also for $n>1$ \cite{macdonald_rezayi}, as is relevant in multi-layer graphene\cite{McCann06, Barlas12,jain221}, the ground states of the resulting Hamiltonian can be characterized as zero energy modes (zero modes). For any $n$, the wave functions of such zero modes will have at least second order zeros as pairs of particles coalesce into the same point. For both $n=1$ and $n=2$, this is equivalent to the polynomial wave function being divisible by the Laughlin-Jastrow factor $\prod_{i<j}(z_i-z_j)^2$.   This was realized early on for $n=1$ \cite{haldane_hierarchy,TK} and leads to the stabilization of the $1/3$-Laughlin-state and its quasi-hole excitations.
The $n=2$ case was extensively discussed by some of us recently \cite{Chen}. The problem of determining the $n=2$ zero mode structure can be considered borderline, as it is still amenable to first-quantized, polynomial methods traditionally employed in single LL problems.
{For $n\geq 3$, zero modes can only be characterized as polynomials belonging to the ideal generated by $(z_i-z_j)^2$ and $(\bar z_i-\bar z_j)^2$ 
for some fixed $i\neq j$, in addition to being anti-symmetric.} This makes the characterization of all possible zero modes considerably more difficult. For the case $n=3$, we will establish that the space of {\em all} zero modes is linearly generated by all wave functions of the form
\begin{equation}\label{ZM}
\psi=\prod_{i<j}(z_i-z_j) \,D_1\,D_2\,,
\end{equation}
where $D_1$ and $D_2$ are the polynomial (in $\{z_i,\bar z_i\}$) parts of two Slater determinants
each comprised of lowest and first excited LL states, and we omit obligatory Gaussian factors.
It is easy to see that states of the form \eqref{ZM} are zero modes of the $n=3$-Hamiltonian. The ``Jain-221'' state where
$D_1=D_2$ is the Slater-determinant of smallest possible angular momentum in the first two Landau levels for given particle number $N$
was conjectured to be the densest zero mode \cite{jain221}. As we will show, the set of all possible wave functions of the form \eqref{ZM} is overcomplete, and we establish rules for the selection of a complete set of zero modes as an EPP on dominance patterns.

\textbf{Entangled Pauli Principle.}
Our starting point shall be a second quantized form of \Eq{eq1} for $n=3$,  in disk geometry, which we present in the general \cite{ortiz}
form
\begin{eqnarray}
{\hat H}_{\sf TK}=\sum_J\sum_{\lambda=1}^{8} E_\lambda \mathcal{T}^{(\lambda)\dagger}_J\mathcal{T}^{(\lambda)}_J.
\label{H2}
\end{eqnarray}
The $\mathcal{T}^{(\lambda)}_J$ annihilate a pair of particles of angular momentum $2J$, with $J=0, \frac 12,1\dotsc$, $\mathcal{T}^{(\lambda)}_J=\sum_{x,m_1,m_2}\eta_{J,x,m_1,m_2}^\lambda c_{m_1,J-x}c_{m_2,J+x}$
and \Eq{H2} may be viewed as a weighted (by $E_\lambda$) sum over eight two-particle projection operators at each $J$. Note that $x$ is (half-odd)-integer if $J$ is (half-odd)-integer, and $c_{m,j}$ destroys a fermion in the $m$th Landau level, $m=0,1,2$, at angular momentum (``site'')
$j\geq -m$. The $\eta$-symbols and the positive $E_\lambda$ can be efficiently derived for general $n$ \cite{mostafaTBP}, and are given for $n=3$ in \cite{SuppMaterial}.
Consider the Slater-determinant decomposition of any $N$-particle zero mode
\be\label{expand}
\ket{\psi}\!=\!\sum C_{m_1,j_1;\dotsc;m_N,j_N}c^\dagger_{m_1,j_1}\dotsc c^\dagger_{m_N,j_N}\ket{0}\equiv \! \sum C_S\ket{S}\,.
\ee
General arguments \cite{ortiz,Chen} imply that there are ``non-expandable'' Slater-determinants $\ket{S}$ in such an expansion
that play a pivotal role in the analysis of any zero mode of \Eq{H2}: These are those states $\ket{S}$ in \Eq{expand} with
non-zero $C_S$ that cannot be obtained from a $\ket{S'}$ with non-zero $C_{S'}$ through an inward-squeezing \cite{bernevigPRL08} process: $\ket{S}\neq  c^\dagger_{m_1,j_1} c^\dagger_{m_2,j_2 }c_{m_2',j_2+x }c_{{m}_1',j_1-x}    \ket{S'}$,
where $j_1<j_2$, $x>0$. We define the state obtained from the zero mode \eqref{expand} by keeping only the non-expandable part
as the ``root state'' $\ket{\psi_{\sf root}}$ of $\ket{\psi}$. The root state is closely related to the thin torus limit \cite{seidel05,seidel2006abelian,bergholtz2006one,bergholtz2006pfaffian, Amila14},
and is generally subject to simple rules usually known as GPPs in the single LL context. We will next show that the zero mode condition leads to a generalization thereof in the present case, which we call EPP.

We begin by showing that a state $\ket{S}$ in  $\ket{\psi_{\sf root}}$ may not have a double occupancy at any given $j$. Otherwise,
we could write $\ket{\psi_{\sf root}}= \sum_{m_1,m_2} \alpha_{m_1m_2} c^\dagger_{m_1,j} c^\dagger_{m_2,j } \ket{\tilde S}+\ket{\sf rest}$, with $\ket{\sf rest}$ being orthogonal to each of the leading terms, and $\ket{\tilde S}$ an $N-2$ particle Slater-determinant with no $j$-mode occupied. The zero mode condition amounts to \cite{ortiz, Chen} $\mathcal{T}^{(\lambda)}_J\ket{\psi}=0$ for all $J$, $\lambda$. Then, in $0=\braket{\psi| {\mathcal{T}^{(\lambda)}_{J=j}}^\dagger|\tilde S}= \sum_{x,m_1,m_2}(\eta_{j,x,m_1,m_2}^\lambda)^\ast  \braket{\psi| c_{m_2,j+x}^\dagger c_{m_1,j-x}^\dagger|\tilde S}$,
the $x\neq 0$ terms must already give zero, otherwise the $x=0$ terms would by definition not appear in $\ket{\psi_{\sf root}}$. One thus obtains the eight conditions
\be
     \sum_{m_1,m_2} \eta_{j,0,m_1,m_2}^\lambda \alpha_{m_1,m_2} =0\;\;(\lambda=1\dotsc 8)\,.
\ee
Since there are only three independent numbers $\alpha_{m_1,m_2}=-\alpha_{m_2,m_1}$, and the $x=0$ $\eta$-symbols are sufficiently \cite{SuppMaterial} linearly independent, one finds that all $\alpha_{m_1.m_2}$ must vanish. One can similarly rule out triple occupancies in $\ket{\psi_{\sf root}}$.
Likewise, one may evaluate possibilities for nearest-neighbor occupancies in $\ket{\psi_{\sf root}}$. Applying the same method to the similar expression ($J$ half-odd integer) $\ket{\psi_{\sf root}}= \sum_{m_1,m_2} \beta_{m_1m_2} c^\dagger_{m_1,J-\frac 12} c^\dagger_{m_2,J+\frac 12 } \ket{\tilde S}+\ket{\sf rest}$, there are eight constraints on the nine constants $\beta_{m_1m_2}$,
\be\label{En1}
     \sum_{m_1,m_2} \eta_{J,1/2,m_1,m_2}^\lambda \beta_{m_1,m_2} =0\;\;(\lambda=1\dotsc 8)\,.
\ee
There is precisely one solution to these equations, which thus {\em determines any nearest-neighbor pair in $\ket{\psi_{\sf root}}$ to be in a certain entangled state}. In evaluating constraints at root level for pairs further separated, we must also take into account inward squeezed configurations of the pair. Writing $\ket{\psi}= \sum_{m_1,m_2} \gamma_{m_1m_2} c^\dagger_{m_1,J-1} c^\dagger_{m_2,J+1 }\ket{\tilde S}+
\alpha_{m_1m_2} c^\dagger_{m_1,J} c^\dagger_{m_2,J } \ket{\tilde S}+ \ket{\sf rest}$, where the first term is non-expandable, we obtain eight conditions in the twelve constants $\gamma_{m_1,m_2}$, $\alpha_{m_1,m_2}=-\alpha_{m_2,m_1}$. After eliminating the latter, these result in five conditions on the $\gamma_{m_1,m_2}$:
\be\label{En2}
     \sum_{m_1,m_2}  \Omega_{J,m_1,m_2}^\mu \gamma_{m_1,m_2} =0\;\;(\mu=1\dotsc 5)\,,
\ee
with $\Omega$ a function of the $\eta$'s at $x=0,1/2$. The constraints derived so far require any two particles in a root state to be entangled when in configurations $\dotsc 11\dotsc$ or $\dotsc 101\dotsc$, where $0$ denotes an empty site, $1$ denotes a single occupancy (in any LL), and consecutive entries denote states with consecutive $j$. We now ask what these constraints imply for clusters of more than two particles.

\textbf{Emergent SU(2)-symmetry.} Let us apply to $\ket{\psi_{\sf root}}$ a non-unitary (but invertible) single particle transformation $\hat V$ such that $\hat V c_{m,j} ^\dagger \hat V^{-1}:=v_{m,s_z}  d_{s_z,j}^\dagger$, where $s_z=0,\pm1$ is interpreted as the SU(2)-label of a spin-1 particle, as detailed in \cite{SuppMaterial}.  In the new basis, \Eq{En1} requires any
nearest-neighbor ($11$) pair in $\hat V\ket{\psi_{\sf root}}$ to form a singlet. Clearly, then, it cannot be entangled with any other particle, which is only consistent with Eqs. \eqref{En1}, \eqref{En2} if any such pair is separated by at least two zeros from any other particle in $\ket{\psi_{\sf root}}$. Moreover, \Eq{En2} takes on a form implying that any $101$-configuration is orthogonal to the spin-2-sector. Studying the satisfiability of this condition for $N$-particles separated by individual empty sites is tantamount to the problem of finding ground states of an open AKLT-chain, leading to the familiar MPS-structure \cite{AKLT}. To label such a structure, we will use the notation
$\dots 1_{\sigma_L}0101\dotsc0101_{\sigma_R}$, where subscripts $\sigma_{L,R}=\pm$ denote the boundary spin-1/2 degrees of freedom of an AKLT ground state. Aside from the aforementioned entangled $11$- and $101$-blocks, a root state may have singly occupied sites surrounded by at least two empty sites on either side. Such sites may be in any of the three LLs, or in any ``spin state'' after the $\hat V$-map.  We denote such configurations by $\dotsc 001_{s_z}00\dotsc$. It follows from all of these observations that a complete set of (rotated) root states can be given by product states of entangled units of the $11$- and $1_{\sigma_L}0\dotsc01_{\sigma_R}$(AKLT)-type,
and of $1_{s_z}$-units, all separated by at least two empty sites. We will refer to the resulting patterns as ``dominance patterns'' compatible with an EPP.

The SU(2)-structure discussed here is not limited to the root level, but {\em emerges} in the full zero mode sector of the Hamiltonian \cite{doi:10.1080/00018730310001642086}.
Indeed, we have identified  {\em non}-Hermitian global SU(2)-generators $S_\nu$, $\nu=x,y,z$ that can be shown to leave the zero mode sub-space invariant. Details are given in \cite{SuppMaterial}. As a consequence, zero-modes can be organized into irreps of this SU(2)-symmetry, as suggested by the root structure and associated dominance patterns.
\begin{table}[t]
\begin{tabular}{c |c}\hline
Patterns&Degeneracy\\\hline
$100...110011001_{s_z}0001_{s_z}$&3$\times$3\\
$100...1100110001_{\sigma_L}01_{\sigma_R}$&4\\
$100...11001_{\sigma_L}0101_{\sigma_R}001_{s_z}$&4$\times$3\\
$100...11001_{\sigma_L}010011_{\sigma_R}$&4\\
$100...1_{\sigma_L}0101010101_{\sigma_R}$&4\\\hline
\end{tabular}
\caption{\label{count3}
Survey of all dominance patterns with angular momentum $\Delta L=3$ above the ground state for odd particle number. The total number including ``spin degeneracy'' allowed by AKLT-entanglement or due to isolated occupied sites is 33, in agreement with Table \ref{counttable}. The corresponding densest state ($\Delta L=0$) has the pattern $100110011\dotsc110011$, where the boundary condition at the left end is explained in \cite{SuppMaterial}.}
\end{table}

\textbf{Braiding statistics.}
Recently, higher LL wave functions have been discussed on the torus \cite{pu2017composite}.
If the dominance patterns established here are understood as ``thin torus (TT) patterns'', there exists a well-defined ``coherent state" method to associate braiding statistics to the excitations of the underlying state \cite{seidel2007domain,seidel08, flavinPRX, Flavin12}.
In this regard, we first observed that if we discard the subscripts $\sigma_{R,L}$ and $s_z$ in the dominance patterns satisfying the EPP,  the resulting reduced patterns of 1s and 0s satisfy the GPP associated with TT/dominance patterns of the $\nu=1/2$ Moore-Read (MR) pfaffian state: There are no more than two 1s in any four adjacent sites. In particular, the densest such patterns, $\dotsc 11001100\dotsc$ and $\dotsc 10101010\dotsc$, signify the six-fold torus degeneracy of the MR-state in the usual way \cite{bergholtz2006pfaffian}. We assume that the EPP developed here will remain meaningful on the torus and govern TT-limits of zero modes of \Eq{eq1}, and that the usual assumptions about adiabatic continuity \cite{seidel05} into the TT-limit hold. Then, in the presence of periodic boundary conditions, the discussion of ground state degeneracy carries over directly from the MR-case, and the torus degeneracy of
the $n=3$ Hamiltonian will be six. However, any charge-1/4 quasi-hole excitation, represented by the familiar domain walls between $1010$ and $1100$-patterns, will carry an additional spin-1/2 described by a $\sigma$-label. So long as we fix the state of this spin (say, $\uparrow$) for all quasi-holes, the coherent state method will make the same predictions for the statistics as in the MR case \cite{seidel08, flavinPRX}.  That is, one finds that each quasi-hole carries a Majorana-fermion, and the operation of braiding two such quasi-holes is described by an operator $\theta_{ij}=\exp(i\theta_m -(-1)^m \frac{\pi}{4} \gamma_i\gamma_j)$, where $\gamma_k$ is the Majorana operator of the $k$th quasi-hole, and $\theta_m$ is a phase only determined up to one of eight possible values by the coherent state method, as reported earlier for the $\nu=1$ bosonic MR-state \cite{seidel08, flavinPRX}. Details for fermions will be given elsewhere, where we show that the method yields $\theta_m=\frac{m\pi}{4}$, $m=0\dotsc 7$. This is consistent with  $\theta=\frac{\pi}{4}$ \cite{nayak} for the $\nu=1/2$ MR-state, but it seems possible that the 221-state discussed here realizes a different one of the allowed phases, which presumably can be determined from the CFT proposed in \cite{wenparton,wen1999projective}. The SU(2)-symmetry discussed above can, however, be used to argue that this phase does not depend on the spin-state of the quasi-holes, and the full braid operator is given simply by $\theta_{ij}X_{ij}$, where $X_{ij}$ exchanges the spin of the $i$th and $j$th quasi-holes.

\textbf{Zero mode counting and edge physics.}
\begin{table}[t]
\begin{tabular}{c |c |c |c |c |c}\hline
$\Delta L$ or $\Delta E$&0&1&2&3&4\\\hline
$N$ odd &1&4&14&33&77\\
$N$ even &3&7&22&50&115\\\hline
\end{tabular}
\caption{\label{counttable} { Number of modes for a  given number of ``quanta'' relative to the ground state. Quanta refers to angular momentum in} the case of microscopic zero modes, and energy in the effective edge theory \eqref{edge}. The counting agrees for at least up to four quanta, and for $\Delta L=3$, is shown in detail in Table \ref{count3} in terms of patterns. The chemical potential in \eqref{edge} is chosen to give equality between total ground state angular momentum and total edge energy for {\em any} particle number, so the agreement also holds in more absolute terms.}
\end{table}
General principles \cite{ortiz,Chen,SuppMaterial} imply that at any given angular momentum $L$, the number of possible dominance patterns
sets an {\em upper} bound on the number of linearly independent zero modes.
This bound has been derived as a
necessary condition on root states (the EPP). As such it applies to a large class of Hamiltonians of the form \Eq{eq1}, and can be generalized to Hamiltonians with different number of terms, internal degrees of freedom, or multi-body interactions. That there are, however, indeed as many zero modes as admitted by the EPP depends strongly on the details of the Hamiltonian. To establish this for the special case of the $n=3$ Hamiltonian \eqref{eq1}, we must show that to each dominance pattern allowed by the EPP, there is a zero mode with the corresponding root state. We show in \cite{SuppMaterial} that indeed, for every dominance pattern one can construct one such
zero mode from the states \eqref{ZM}. This then necessarily yields a complete set of zero modes. In particular, it is easy to show that the Jain-221
state has $\ket{\psi_{\sf root}}$ corresponding to the densest possible pattern consistent with the EPP: $\dotsc 11001100110011\dotsc$.
This rigorously establishes that the Jain-221 state is the densest possible zero mode, since there are no allowed dominance patterns
at higher filling factor, or smaller $L$ at given $N$. The existence of a densest filling factor (here: 1/2) permitting zero modes is usually taken as  indication for incompressibility at this filling factor. This is in particular so if the edge theory encoded in the zero mode counting is a unitary rational conformal field theory (CFT). Via patterns, we now have full control over zero mode counting. Let ${\cal N}(\Delta L)$ be the number of zero modes of \Eq{eq1} at angular momentum $\Delta L$ relative to the ground state, where $\Delta L \ll N$.
One may ask \cite{milovanovich, read08} if ${\cal N}(\Delta L)$ so defined agrees with the number of states  having $\Delta L$ energy quanta in some CFT. Indeed, in the presence of suitable chemical potential terms, one may find \cite{Chen} complete agreement, for $\Delta L \ll N$, between the degeneracies of some CFT Hamiltonian and of the  {\em total} angular momentum operator $\hat L$ within the zero mode sector of a special Hamiltonian, for any fixed particle number $N$ ($N$ being identified with a suitable conserved quantity of the CFT). For $\Delta L\leq 4$, we verified such agreement between the mode counting determined by the our EPP  and the mode counting in a 1+1d edge theory of the form \cite{wenparton,wen1999projective}
\be\label{edge}
   H=\sum_{i=0,1} H_{b,i}(\Phi_i)+H_f(\gamma) -\frac 52 N_0\,,
\ee
where the $\Phi_i$ are free chiral bosons of compactification radii $\frac 12$ and $1$, respectively, $\gamma$ is a Majorana field in the anti-periodic sector, all modes are co-propagating, $N_i$ is the winding number of $\Phi_i$, and the parity of the number of occupied Majorana modes must be opposite to $N_0+N_1$.
Except for the chemical potential term, \Eq{edge} is the $U(1)\times SU(2)_2$ edge CFT first ascribed to the Jain-221 state in Refs. \cite{wenparton,wen1999projective}.
When $N_0$ is identified with the particle number $N$, the mode-counting agreement described above holds as summarized in Table \ref{counttable} and caption. Detailed counting for the number of zero modes at $\Delta L=3$ in terms of patterns is shown in Table \ref{count3}.

\textbf{Conclusion.}
The formalism presented here gives controlled access to a large class of quasi-exactly solvable quantum-many-body Hamiltonians and combines features of earlier studied cases with the added complexity of LL mixing.
We demonstrated that entanglement present at root level generalizes previously known organization principles for zero mode spaces. We call this emergent phenomenon Entangled Pauli Principles (EPPs). As an application, we microscopically verified field-theoretic predictions about the Jain-221 state in detail.
Specifically, within the coherent-state framework that utilizes dominance patterns as an input \cite{flavinPRX}, non-Abelian braiding statistics can be predicted. Moreover, through rigorous organization principles of all the zero modes of a microscopic parent Hamiltonian, in particular the edge physics can be fully characterized.  This confirms a picture \cite{wenparton, wen1999projective, overbosch} of three co-propagating edge modes consisting of two bosonic branches and one Majorana fermion, making this state distinctly different from other non-Abelian candidate states for a half-filled Landau level, notably the MR pfaffian \cite{MR} and the anti-Pfaffian \cite{levinAP,leeAP} states. In its bulk, the Jain-221 state is further distinct from these other candidates by the presence of two types of charge 1/4 quasi-particles.
All this is efficiently captured by the EPP, which for the most part eliminates the LL degrees of freedom from mode counting, except by dressing up domain walls with local pseudo-spin degrees of freedom. The topological shift on the sphere, which further distinguishes candidate states at $\nu=1/2$ and in principle relates to Hall viscosity \cite{Read09, RR10}, is likewise efficiently encoded by these patterns.
The present work unambiguously establishes the emergence of non-Abelian phases in quasi-exactly solvable {\em two}-body Hamiltonians, which may be an attractive prospect, e.g., for realization in cold atom systems, or for the study of spectral bounds as discussed in \cite{Amila15}. The 3-fold LL degeneracy of our model may further lead to intriguing consequences for ABC stacked trilayer graphene \cite{jain221, McCann06, Barlas12}. 
We leave the exploration of larger $n$, { different types of interactions}, and bosons as interesting problems for the future.

\begin{acknowledgments}
The authors thank J. Jain and K. Yang for insightful discussions.
LC's work was supported in part by Department of Energy, Office of Basic Energy Sciences through Grant No. DE-SC0002140, and performed at the National High Magnetic Field Laboratory, which is supported by National Science Foundation Cooperative Agreements No. DMR-1157490 and No. DMR-1644779, and the State of Florida.
\end{acknowledgments}

\bibliography{Nat0}
\onecolumngrid
\vspace{\columnsep}
\begin{center}
\begin{Huge}
Supplemental Material
\end{Huge}
\end{center}
\vspace{\columnsep}
\twocolumngrid
\section{Second quantization in disk geometry\label{3}}
We will present a general method to project $H_{\sf TK}$ onto the lowest $N_L$ Landau levels \cite{mostafaTBP},
specializing to the $N_L=3$ case.

Since we want to project a two-body Hamiltonian, we  construct an appropriate two-fermion basis
\begin{eqnarray}
\Phi_{JI}&=&G_{(-1)^{m+1}}^{n_1,n_2} \Phi_{0[J+(n_1+n_2)/2]}^m ,
\label{defofstate}
\end{eqnarray}
where, $\Phi_{0J}^m$ is a lowest Landau level state of two particles with relative angular momentum $m$ and total angular momentum $2J$,
\begin{eqnarray}\hspace*{-0.3cm}
\Phi_{0J}^m=\frac{2^{-J}}{\sqrt{(2J-m)! \, m!}} \
(b_1^\dagger+b_2^\dagger)^{ 2J-m} (b_1^\dagger-b_2^\dagger)^{ m} \Phi_0 ,
\label{vac}
\end{eqnarray}
elevated to higher Landau levels by the operator $G_{\pm}^{n_1,n_2}$ with $0\leq n_i \leq N_L-1$ and,
\begin{eqnarray} \hspace*{-0.5cm}
G_{\pm}^{n_1,n_2}\!&=&\! \frac{1}{\sqrt{n_1!n_2! \
2(1+\delta_{n_1,n_2})}}(a^{\dagger n_1}_1a^{\dagger n_2}_2
\pm a^{\dagger n_2}_1a^{\dagger n_1}_2).
\label{Goperator}
\end{eqnarray}
$I$ encodes a multi-index consisting of the quantum numbers $n_1$, $n_2$, and $m$ as per Table \ref{triplet}, and $\Phi_0$ is the two-particle vacuum of the ladder operators $a_{1,2}$, $b_{1,2}$ associated to dynamical momenta and guiding centers, respectively, which can be defined in symmetric gauge as
\begin{eqnarray}
a_i^{\;}=\frac{1}{\sqrt{2}}(\frac{z_i}{2\ell}+2\ell \partial_{\bar{z}_i}) \ , \
a_i^{\dagger}=\frac{1}{\sqrt{2}}(\frac{\bar{z}_i}{2\ell}-2\ell\partial_{z_i}) ,
\end{eqnarray}
\begin{eqnarray}
b_i^{\;}=\frac{1}{\sqrt{2}}(\frac{\bar{z}_i}{2\ell}+2\ell\partial_{z_i}) \ , \
b_i^{\dagger}=\frac{1}{\sqrt{2}}(\frac{z_i}{2\ell}-2\ell\partial_{\bar{z}_i}),
\end{eqnarray}
$i=1,2$. The latter satisfy the canonical bosonic algebra
\begin{eqnarray}
[a_i^{\;},a_j^{\dagger}]=\delta_{ij}=[b_i^{\;},b_j^{\dagger}] \ , \  [a^{\;}_i,b^{\;}_j]=[a^\dagger_i,b_j^{\dagger}]=0 .
\end{eqnarray}
Note that  Eq. (\ref{defofstate}) is even (odd) in $n_1$, $n_2$ for $m$ odd (even),
thereby always producing a state that's odd under the exchange of particle coordinates.

 We are interested in establishing the Fock-space representation of $H_{\sf TK}$ projected onto
 the subspace of the three lowest Landau levels, $0\leq n_i\leq 2$, generated by the basis $\Phi_{JI}$. Note that the latter are orthogonal by construction. It further turns out that $H_{\sf TK}$ annihilates all states with $m>n_1+n_2+1$. For any fixed $J$, its nonzero eigenvalues and eigenstates can therefore be obtained by diagonalizing $H_{\sf TK}$ within the subspace defined by the 18 $I$-indices listed in Table \ref{triplet}. Moreover, the relevant matrix elements can be shown to be {\em independent} of $J$. We will thus omit the $J$-index from now on when no confusion may arise. The 18-dimensional subspace defined in Table \ref{triplet} contains all positive eigenvalue eigenstates for two particles at given $J$. Straightforward but tedious diagonalization yields that there are only eight such states (with all orthogonal states, even within this subspace, having zero energy), as listed in Table \ref{tt1}.
\begin{table}[t]
\begin{tabular}{c|cccccccccccccccccc}\hline\hline
 I & 1 & 2 & 3 & 4 & 5 & 6 & 7 & 8 & 9 & 10 & 11 & 12 & 13 & 14 & 15 & 16 & 17 & 18 \\\hline
 $n_1$ & 0 & 0 & 1 & 1 & 1 & 1 & 2 & 2 & 2 & 2 & 2 & 2 & 2 & 2 & 2 & 2 & 2 & 2 \\
 $n_2$ & 0 & 1 & 0 & 0 & 1 & 1 & 0 & 0 & 0 & 0 & 1 & 1 & 1 & 1 & 1 & 2 & 2 & 2 \\
 $m$ & 1 & 0 & 2 & 1 & 1 & 3 & 1 & 3 & 0 & 2 & 1 & 3 & 0 & 2 & 4 & 1 & 3 & 5 \\\hline\hline
\end{tabular}
\caption{Triplets $(n_1, n_2, m)$ for any given state $\Phi_I$ with $I=1,2,..., 18.$}
\label{triplet}
\end{table}
We formally write these eigenstates as
 \begin{eqnarray}\label{expansion}
 \Psi_\lambda=\sum_{I}\alpha^\lambda_I \Phi_I,
 \end{eqnarray}
 with $\lambda$ an index associated to the eight positive eigenvalues $E_\lambda$ and  coefficients $\alpha^\lambda_I$ made explicit in Table \ref{tt1}. Passing to a second quantized language is now easy. We write the two-particle states \eqref{defofstate} as
 $\ket{\Phi_{JI}}=T^\dagger_{JI}\ket{0}$, with $\ket{0}$ the vacuum of the Fock space. The two-particle creation operators $T^\dagger_{JI}$
 can be written as \cite{mostafaTBP}
 \begin{eqnarray}\label{TJI}
T_{JI}^{\dagger}=\frac{1}{\sqrt{2(1+\delta_{n_1,n_2})}}
&&\sum_{k}\eta_{k+\frac{n_2-n_1}{2}}(J+\frac{n_1+n_2}{2},m_I)\nonumber\\
&&\,c_{n_1,J-k}^{\dagger} c_{n_2,J+k}^{\dagger},
\end{eqnarray}
where $-J-n_2\leq k \leq J+n_1$ for
the infinite plane\footnote{
For a finite size disk with $L$ available states, the last inequality must be replaced with
$-\text{min}(J,L-1-J)\leq k \leq \text{min}(J,L-1-J)$.}.
In \Eq{TJI}, $c^\dagger_{n,x}$ creates an electron in LL $n$ with angular momentum $x$, and
the form factor $\eta_k(J,m)$ is the identical to the one already appearing in the lowest Landau level case \cite{ortiz},
\begin{eqnarray}
\eta_p(J,m)&=&2^{-J+1/2}\sqrt{\frac{(J-p)!\,(J+p)!}{(2J-m)!\,m!}}(-1)^{m+J-p}\nonumber \\
&&\times\sum_{r=0}^{J-p}(-1)^{r}\binom{2J-m}{r}\binom{m}{J-p-r}.\nonumber\\
\end{eqnarray}

One can write the states of \Eq{expansion} as
 \begin{eqnarray}
 |\lambda \rangle_J=\sum_I\alpha^\lambda_I \,T_{JI}^{\dagger} \ket{0}= \mathcal{T}^{(\lambda)\dagger}_J\ket{0},\label{ketlambda}
 \end{eqnarray}
where
\begin{eqnarray}
\mathcal{T}_J^{(\lambda)}&=&\sum_{I,k}\frac{\alpha_I^\lambda}{\sqrt{2(1+\delta_{n_1,n_2})}}
\eta_{k+\frac{n_2-n_1}{2}}(J+\frac{n_2+n_1}{2},m_I)\nonumber\\
&&\,c_{n_{2},J+k} c_{n_{1},J-k}\nonumber\\
&:=&\sum_{k, n_1,n_2}
\eta^\lambda_{J,k,n_1,n_2}
\,c_{n_2,J+k} c_{n_1,J-k}\nonumber\\
\end{eqnarray}
and we have made contact with the $\eta^\lambda$-symbols defined in the main text, letting
\begin{eqnarray}\label{etatrans}
\eta^\lambda_{J+\frac{n_2+n_1}{2},k,n_1,n_2}=
&&\sum_{I}\frac{\alpha_I^\lambda}{\sqrt{2(1+\delta_{n_1,n_2})}}\nonumber\\
&&\eta_{k+\frac{n_2-n_1}{2}}(J+\frac{n_2+n_1}{2},m_I).
\end{eqnarray}

Now the two-particle Hamiltonian can now be written manifestly
in terms of its
spectral decomposition,
\begin{equation}
H_{\sf TK}= \sum_J \sum_{\lambda=1}^8 E_\lambda ~_J|\lambda\rangle\langle\lambda|_J,
\end{equation}
and generalized, as usual, to a many-body Hamiltonian by dropping the projection $|0\rangle\langle 0|$ onto the vacuum that, upon use of \Eq{ketlambda}, would otherwise follow  the action of $\mathcal{T}^{(\lambda)}_J$:
\begin{eqnarray}
{\hat H}_{\sf TK}=\sum_J\sum_{\lambda=1}^{8} E_\lambda\, \mathcal{T}^{(\lambda)\dagger}_J\mathcal{T}^{(\lambda)}_J.
\label{Ham}
\end{eqnarray}

\begin{table*} [!b]

\large

\begin{ruledtabular}
\begin{tabular}{c|c}
$4 \pi E_\lambda$ &Eigenvectors \tabularnewline
\hline
$\frac{87}{8}$ & $\Psi_1=
2 \sqrt{\frac{2}{15}}\Phi_1 -\frac{4}{\sqrt{15}}\Phi_3 -\frac{2}{\sqrt{
  5}}\Phi_6+ \frac{2}{\sqrt{5}}\Phi_8
+ 2 \sqrt{\frac{2}{5}}\Phi_{15}+ \Phi_{18}$  \\ \tabularnewline
$\frac{3}{4}(6+\sqrt{17})$ & $\Psi_2=
 -\frac{\sqrt{6}(363+89 \sqrt{17})}{3 (469+113 \sqrt{17})}\Phi_2 -\sqrt{\frac{2}{3}}\Phi_5
  +\frac{\sqrt{6} \left(7 \sqrt{17}+23\right)}{33 \sqrt{17}+141}\Phi_7
 +\frac{31 \sqrt{17}+129}{\sqrt{6} \left(11 \sqrt{17}+47\right)}\Phi_{14}+\Phi_{17}$ \\ \tabularnewline
$\frac{15}{4}$ & $\Psi_3=-\frac{2}{\sqrt{3}}\Phi_4+
 2 \sqrt{\frac{2}{3}}\Phi_{10}+ \Phi_{12}$ \\ \tabularnewline
$\frac{1}{8}(9+\sqrt{33})$ & $\Psi_4=
  \Phi_9+ \frac{1+\sqrt{33}}{4 \sqrt{2}}\Phi_{11}$ \\ \tabularnewline
$\frac{3}{4}(6-\sqrt{17})$ & $\Psi_5=
\frac{\sqrt{6} \left(89 \sqrt{17}-363\right)}{3 \left(469-113 \sqrt{17}\right)}\Phi_2-\sqrt{\frac{2}{3}}\Phi_5
 +\frac{\sqrt{6} \left(7 \sqrt{17}-23\right)}{33 \sqrt{17}-141}\Phi_7
 +\frac{31 \sqrt{17}-129}{\sqrt{6} \left(11 \sqrt{17}-47\right)}\Phi_{14}+\Phi_{17}$ \\  \tabularnewline
$\frac{1}{16}(9+\sqrt{57})$ & $\Psi_6=
 \Phi_{13}+\frac{5+\sqrt{57}}{4 \sqrt{2}}\Phi_{16}$ \\  \tabularnewline
$\frac{1}{8}(9-\sqrt{33})$ & $\Psi_7=
  \Phi_9+ \frac{1-\sqrt{33}}{4 \sqrt{2}}\Phi_{11}$ \\ \tabularnewline
$\frac{1}{16}(9-\sqrt{57})$ & $\Psi_8=
 \Phi_{13}+\frac{5-\sqrt{57}}{4 \sqrt{2}}\Phi_{16}$
\tabularnewline \\
\end{tabular}
\caption{Eigenvalues and Eigenvectors of the projected $H_{\sf TK}$. Overall normalization factors in the column to the right are omitted. They are straightforward but tedious to calculate, and are not needed throughout the paper.}
\label{3LL}

\end{ruledtabular}\label{tt1}
\end{table*}

\section{Construction of EPP from microscopic Hamiltonian\label{4}}

In this section, we provide some additional details for the derivation of the EPP from the second-quantized zero mode condition associated with the microscopic Hamiltonian derived in the preceding section.
We reproduce this zero mode condition here as

\be
\mathcal{T}_J^{(\lambda)} \ket{\psi}=0\quad \forall \,J,\lambda.
\ee
Note that equivalent reformulations of these conditions can be given in terms of arbitrary new (linearly independent) linear combinations of the
$\mathcal{T}_J^{(\lambda)}$. From Table \ref{tt1}, is easy to see that the $T_{JI}$ with $I=9,11,13,16$ must all individually annihilate any zero mode. Moreover, from $\mathcal{T}_J^{(2)}$ and $\mathcal{T}_J^{(5)}$, we may
make new linear combinations
\begin{eqnarray}
 \tilde{\mathcal{T}}
 _J^{(2)}&=& T_{J,2}-2T_{J,7}-\frac 12 T_{J,14},\nonumber\\
  \tilde {\mathcal{T}}_J^{(5)}&=& T_{J,5}+T_{J,7}- T_{J,14}-\sqrt{\frac 32} T_{J,17},
\end{eqnarray}
so that we may rephrase the zero mode condition for a ket $\ket{\psi}$ equivalently by saying that $\ket{\psi}$ is annihilated by each of the eight operators in the set $Z_J=\{{\mathcal{T}}
 _J^{(1)}, \tilde{\mathcal{T}}
 _J^{(2)}, {\mathcal{T}}
 _J^{(3)}, \tilde{\mathcal{T}}
 _J^{(5)}, T_{J,9}, T_{J,11}, T_{J,13}, T_{J,16}\}$, for all $J$. This considerably simplifies the resulting equations.

We first turn to Eq. (5) of the main text, which we rephrase here for the operators in the set $Z_J$:
\be\label{20eq}
     \sum_{n_1,n_2} \eta_{J,0,n_1,n_2}^{(\tilde{\lambda})} \alpha_{n_1,n_2} =0\;\;(\lambda=1\dotsc 8)\,.
\ee
where $\tilde{\lambda}$ now indexes the members of the set $Z_J$, and $\eta^{(\tilde{\lambda})}$ is the associated form factor.
The goal is to show that these have only trivial solutions.
Since there are only three independent variables $\alpha_{n_1,n_2}=-\alpha_{n_2,n_1}$, it is sufficient to focus on three members of $Z_J$.
The $T_{J,9}$-equation in \eqref{20eq}  then readily implies $\alpha_{02}=0$ (cf. Table \ref{triplet}), and the $T_{J,13}$-equation implies $\alpha_{12}=0$.
Finally, consider the $\mathcal{T}_{J}^{(2)}$-equation.
Since $\eta_0(J,m)=0$ in \Eq{TJI} for $m$ odd, the only contributions to this equation can come from $T_{J,2}$ and $T_{J,14}$ (Tables \ref{triplet} and \ref{tt1}). However, that of  $T_{J,14}$ also vanishes, since $\alpha_{12}=0$ is already known. This gives $\alpha_{01}=0$.

We may likewise put Eq. (6) of the main text into a form that references the form factors associated to the operator set $Z_J$:
\be\label{21eq}
     \sum_{n_1,n_2} \eta_{J,1/2,n_1,n_2}^{(\tilde{\lambda})} \beta_{n_1,n_2} =0\;\;(\lambda=1\dotsc 8)\,.
\ee
The resulting eight linear equations have the following solution, unique up to a scale factor:
\begin{equation}
\begin{split}\label{betas}
\beta_{22}=\beta_{12}=\beta_{21}=\beta_{10}=0, \beta_{20}=1, \beta_{11}=-\sqrt{2},\\ \beta_{01}=\frac{\sqrt[]{8}}{\sqrt[]{J+1}}, \beta_{02}=\frac{\sqrt[]{J+3}}{\sqrt[]{J+1}},  \beta_{00}=\frac{\sqrt[]{2(J+2)}}{\sqrt[]{J+1}}.
\end{split}
\end{equation}
At root level, as explained in the main text, this uniquely fixes any nearest neighbor occupied orbitals to be in a certain entangled state. Upon the local change of basis detailed in the next section, we can understand this state as a ``singlet'' formed by two spin-1 degrees of freedom. In the dominance patterns that we use to encode root states, this two-orbital entangled state is simply represented as $\dotsc 11\dotsc$.

Last, we also consider the situation of occupied next-nearest neighbor orbitals in some more detail. As in the main text, consider a zero mode of the form $\ket{\psi}= \sum_{n_1,n_2} \gamma_{n_1n_2} c^\dagger_{n_1,J-1} c^\dagger_{n_2,J+1 }\ket{\tilde S}+
\alpha_{n_1n_2} c^\dagger_{n_1,J} c^\dagger_{n_2,J } \ket{\tilde S}+ \ket{\sf rest}$, where $\ket{\tilde S}$ is an $N-2$ particle Slater-determinant that has all orbitals with angular momenta $J$, $J\pm 1 $ vacant, $\ket{\sf rest}$ is orthogonal to the first two terms, and the first term is non-expandable.
The condition $0=\braket{\psi| {\mathcal{T}^{(\tilde\lambda)}_{J}}^\dagger|\tilde S}= \sum_{x,n_1,n_2}(\eta_{J,x,n_1,n_2}^{(\tilde \lambda)})^\ast  \braket{\psi| c_{n_2,j+x}^\dagger c_{n_1,j-x}^\dagger|\tilde S}$ then leads to the conditions
\begin{equation}
\sum_{n_1,n_2}
\left(\eta_{J,1,m_1,m_2}^{(\tilde\lambda)}
\gamma_{n_1,n_2} + \eta_{J,0,m_1,m_2}^{(\tilde\lambda)}
\alpha_{n_1,n_2} \right)=0\,,
\end{equation}
where again only the $x=0$ and $x=1$ terms can contribute, as the presence of any other terms would imply that the $\gamma_{n_1,n_2}$-terms could be obtained via inward squeezing, contrary to assumption. From these eight equations, the three variables $\alpha_{n_1,n_2}=-\alpha_{n_2,n_1}$
may be eliminated, leaving five equations for the coefficients
$\gamma_{n_1,n_2}$ that constrain the entanglement of second-nearest neighbor occupied orbitals at root level:
\begin{widetext}
\begin{small}
\begin{eqnarray}\label{gammas}\nonumber
&&\gamma_{22}=0,~~~\gamma_{21}+\sqrt[]{\frac{2+J}{4+J}}\gamma_{12}=0,
~~~\gamma_{00}-\frac{2}{\sqrt[]{3+J}}\gamma_{01}+\sqrt{\frac{1+J}{3+J}}\gamma_{11}+\sqrt{\frac{18}{(3+J)(4+J)}}\gamma_{02}-\sqrt[]{\frac{8(1+J)}{(3+J)(4+J)}}\gamma_{12}=0\,,\\
&&\sqrt[]{3+J}\gamma_{10}+\sqrt[]{1+J}\gamma_{01}-\sqrt[]{\frac{2}{(4+J)}}\gamma_{12}-\sqrt[]{\frac{8(1+J)}{(4+J)}}\gamma_{02}=0,~~~\gamma_{20}+\sqrt[]{\frac{2+J}{3+J}}\left(\gamma_{11}-\frac{2}{\sqrt[]{4+J}}\gamma_{12}-\sqrt[]{\frac{1+J}{4+J}}\gamma_{02}\right)=0\,.
\end{eqnarray}
\end{small}
\end{widetext}
The intuitive meaning of these equations will again become clearer in the following section. There are four solutions to these five equations in nine variables, which we formally label
as $1_\uparrow 0 1_\uparrow$, $1_\uparrow 0 1_\downarrow$, $1_\downarrow 0 1_\uparrow$, and
$1_\downarrow 0 1_\downarrow$. A dominance pattern containing one of these strings $\dotsc00 1_{\sigma_1}01_{\sigma_2}00\dotsc$ corresponds to a root state where the two orbitals indicated by the $1$'s in the pattern are in a pure entangled state corresponding to one of the four solutions. Conversely, in any root state of a zero mode, the state of any two next-nearest-neighbor occupied orbitals must always be in the four-dimensional subspace defined by these four solutions. As long as no member of the pair has any other nearest or next-nearest neighbor orbitals occupied, there are no further constraints affecting the pair. However, if one member had a nearest neighbor occupied, as in the string $1011$, equations \eqref{gammas} constraint the first pair, while equations \eqref{betas} constrain the second. There are no solutions to the combined set of equations, thus there are no dominance patterns of the $1011$ kind. Similarly, the string $111$ can be ruled out, and a $11$ configuration must thus always be separated by $00$ on either side from all the other orbitals, in any legitimate dominance pattern. The only remaining case of interest is that of consecutive strings of next nearest neighbors. In such strings, Eqs. \eqref{gammas} must be applied to each next-nearest-neighbor pair. We will see in the next section that the resulting equations, applied to any string of consecutively occupied next-nearest-neighbor orbitals separated by terminal $00$ units from all other orbitals, still result in four solutions. We will show this below by showing that solutions have an MPS-structure that's of a kind with ground states in the AKLT model. The resulting dominance patterns are thus again naturally labeled by strings
$\dotsc001_{\sigma_1}0101\dotsc 10101_{\sigma_2}00\dotsc$, where only the terminal $1$s carry a spin-1/2 index labeling a boundary degree of freedom.
In all, we have shown that states appearing at root level for any zero mode can be decomposed into mutually non-entangled units of the following kinds: 1. Nearest neighbor pairs $11$ governed by Eqs. \eqref{betas}, 2. next-nearest neighbor strings $1_{\sigma_1}01\dotsc 101_{\sigma_2}$, and 3. isolated occupied sites $1_{s_z}$, where $s_z$ may be interpreted either as a label for the three Landau levels or, alternatively, a spin-1 label to be discussed in the following section.
All these units must be separated by at least two unoccupied sites from one another. Special consideration must be given to the orbitals with negative angular momenta $j=-1$ and $j=-2$. Carrying out the above analysis  with the special constraint in mind that there is only one such orbital for $j=-2$ and two such orbitals for $j=-1$, one obtains the boundary condition that at root level, apart from being unoccupied, the $j=-2$ orbital may only occupy isolated $1_{s_z={\sf max}}$ unit. Again, the latter must again be separated by at least two zeros from all other units.
Similarly, the $j=-1$ orbital may
only be in a $1_{s_z}$ state, with $s_z$ assuming the top two values, or may be the left end of a $1_{\sigma_L}0101\dots$ pattern with $\sigma_L$ fixed to $\uparrow$. 
This completes the set of rules that all dominance patterns and their associated root states are subject to.

We emphasize that thus far, the above rules represent necessary conditions
on root states. Below we establish that to each permissible dominance pattern, there is precisely one zero mode that has the associated root state. Since zero modes form a linear space, the root state of a generic zero mode may, of course, as well be a superposition of root states associated with the dominance patterns characterized above.

\section{Emergent SU(2)-Symmetry }

We now discuss an emergent SU(2)-symmetry within the zero mode sector that also sheds the entangled Pauli principle discussed in the proceeding section in a simpler light. To this end, we temporarily limit the discussion to the Fock space ${\cal F}_+$ associated to orbitals of angular momentum index $j\geq 0$ (and, as before, LL index $0\leq n\leq 2$). We consider the following single particle operators acting within this space, which we define in first quantization through their action on the {\em polynomial part} of the wave function via
\begin{equation}
\begin{split}
S_z&=\sum_{i}(\bar{z}_i\partial_{\bar{z}_i}-1),\\
S_-&=\sum_{i}\frac{1}{z_i}\partial_{\bar{z_i}},~~~S_+=\sum_{i}z_i\bar{z}_i(2-\bar{z}_i\partial_{\bar{z}_i}).\label{su2}
\end{split}
\end{equation}
If the action on full wave functions, including Gaussian factors, is desired, a simple shift
$\partial_{\bar z_i}\rightarrow \partial_{\bar z_i}+\frac 14 z_i$ may be performed. In this section, we will omit Gaussian factors for simplicity.

One checks without difficulty that
the operators \eqref{su2} satisfy the su(2)-algebra $[S_+,S_-]=2S_z$,
$[S_z,S_\pm]=\pm S_\pm$, albeit without having the properties under Hermitian conjugation that are usually taken for granted in physics. This is irrelevant to the representation theory of this algebra, and in any case the representation within ${\cal F}_+$
can be unitarized by using the following single particle basis:
\begin{equation}\label{mono}
z^j,\;\;\sqrt{2}z^{j+1}\bar z, \;\;
z^{j+2}\bar z^2\,.
\end{equation}
In this basis, it is manifest that each angular momentum $j\geq 0$ is associated to a triplet of LL orbitals that transforms under the spin-1 representation of the operators \eqref{su2}. The usual Landau level basis is obtained by applying to the above, written as a column vector, the matrix
\small{\begin{equation}
V=\frac{1}{\sqrt[]{2\pi2^jj!}}\left(
\begin{array}{ccc}
 1 & 0 & 0 \\
 -\sqrt{j+1} & \frac{1}{2 \sqrt{j+1}} & 0 \\
 \frac{\sqrt{(j+1) (j+2)}}{\sqrt{2}} & -\frac{\sqrt{j+2}}{\sqrt{2} \sqrt{(j+1) }} & \frac{1}{4 \sqrt{2} \sqrt{(j+1) (j+2)}} \\
\end{array}
\right),
\end{equation}}
whose matrix elements
$v_{m,s_z}$ are referenced in the main text to define operators $d_{s_z,j}^\dagger$. The latter just create the single particle states \eqref{mono}. From \Eq{mono} it is also clear that the space ${\cal F}_+$ is invariant under the action of the generators \eqref{su2}.
If we define ${\cal F}^0_+$ as the subspace of zero modes that are contained in ${\cal F}_+$, we want to show next that ${\cal F}^0_+$ is {\em also} invariant under the action of the generators. These operators thus generate an {\em emergent} (since the Hamiltonian is not invariant) symmetry within the zero mode subspace ${\cal F}^0_+$.

It is sufficient to analyze this question for two-body wave functions.
Take $S_-$ and act on a two body wave function in ${\cal F}_+^0$, which we  express as a polynomial $\psi(Z, \bar Z, z, \bar z)$ in the center-of-mass and relative coordinates $Z=\frac 12 (z_1+z_2)$, $z=z_1-z_2$ and their complex conjugates. Being a zero mode, $\psi$ has a third order zero in $z$, $\bar z$ for any $Z$, $\bar Z$. Moreover, since $S_-$ certainly preserves analyticity for $|z|<2|Z|$, $|\bar z|<2|\bar Z|$, and contains only single derivatives, $S_-\psi$ must still have at least a second order zero in $z$, $\bar z$ for any $Z, \bar Z\neq 0$. As $S_-$ also preserves oddness under $z\rightarrow -z$, $\bar z\rightarrow -\bar z$, $S_-\psi$ must in fact {\em still} have  a third order zero in $z$, $\bar z$ for any $Z, \bar Z\neq 0$. On the other hand, since $\psi \in {\cal F}_+$, $S_-\psi$ is still in ${\cal F}_+$, and is still analytic everywhere (in fact polynomial). If in its expansion
\begin{equation}
   S_- \psi = \sum_{m,n\geq 0} z^n\bar z ^m g_{mn}(Z,\bar Z)
\end{equation}
there is any non-zero term with $n+m<3$, then $g_{m,n}(Z,\bar Z)$ is a polynomial of non-zero degree and must be finite at some $Z, \bar Z\neq 0$. At such $Z,\bar Z$, $S_- \psi$ would then {\em not} have a third order zero in $z$, $\bar z$, contradicting the foregoing. Therefore, all $g_{mn}$ with $m+n<3$ vanish, and $S_-\psi$ is in ${\cal F}_+^0$. The cases $S_z$ and $S_+$ can be treated similarly (and without paying special attention to $Z,\bar Z=0$). ${\cal F}_+^0$ is thus invariant under the generators \eqref{su2}.

We emphasize that the notion of an emergent SU(2) symmetry is not an artifact of the restriction  to ${\cal F}_+$. Note that any zero mode of well-defined total angular momentum (thus finite spatial extent) will, up to exponentially small terms, lie in ${\cal F}^0_+$ after a sufficiently large spatial translation $T$. Action with the modified generators $\tilde S_i=T^\dagger S_i T$ will preserve the zero mode property, up to terms that can be made exponentially small. Note that the $\tilde S_i$ are still {\em local} operators (though no longer angular momentum preserving). Related to that, the construction of the generators \eqref{su2} naturally extends to the cylinder geometry. There, the singularity at $z_i=0$  (for the disk geometry) is automatically pushed to infinity.

The single particle orbitals \eqref{mono} may be extended to $j\geq -2$, with the additional constraint that orbitals with negative exponents are to be discarded. The resulting set of orbitals is then a non-orthogonal basis of the three lowest landau levels (associated to the $d_{s_z,j}^\dagger$ operators of the main text, where $s_z\geq \max(-1,-1-j)$). It is natural to analyze the conditions \eqref{betas} and \eqref{gammas} in this basis. It is straightforward to show that \Eq{betas} precisely expresses that any $11$ factor of a root state must be a singlet under the su(2) algebra \eqref{su2}. Moreover, \Eq{gammas}
mandates that any neighboring particles in a $\dotsc 101\dotsc $ factors must have total spin 0 or spin 1 (i.e., after introduction of an inner product for which the orbitals \eqref{mono} are orthonormal, any $101$ in a root state must be orthogonal to spin 2). This is precisely the zero mode condition of the famous AKLT-model \cite{AKLT}.
The claims about the MPS-structure and number of solutions to the constraints \eqref{gammas} made above and in the main text are immediate consequences of this observation.

\Comment{For future reference, we express the operators \eqref{su2} in  second quantization:

\be \begin{split}
S_z= &\sum_{r\ge 0}(-c_{0,r}^\dag c_{0,r}+\sqrt{r+1}c_{0,r}^\dag c_{1,r}\\&+ c_{2,r}^\dag c_{2,r}+\sqrt{2(r+2)}c_{1,r}^\dag c_{2,r}),
\end{split}\ee
\be \begin{split}
S_-= &\sum_{r\ge 0}\Big(\frac{1}{2}\sqrt{\frac{1}{r+1}}c_{0,r}^\dag c_{1,r}+\sqrt{\frac{1}{2(r+2)}}c_{1,r}^\dag c_{2,r}\\&-\sqrt{\frac{1}{2(r+1)(r+2)}}c_{0,r}^\dag c_{2,r}\Big),
\end{split}\ee
\be \begin{split}
S_+= &\sum_{r\ge 0}(4\sqrt{r+1}c_{1,r}^\dag c_{0,r}+4(r+1)c_{0,r}^\dag c_{0,r}\\&+2\sqrt{2(r+2)}c_{2,r}^\dag c_{1,r}+4 c_{1,r}^\dag c_{1,r} \\&-2r\sqrt{r+1}c_{0,r}^\dag c_{1,r}-4(r+2) c_{2,r}^\dag c_{2,r}\\&-2(r+3)\sqrt{2(r+2)}c_{1,r}^\dag c_{2,r}\\&-2\sqrt{2(r+1)(r+2)}c_{0,r}^\dag c_{2,r}).
\end{split}\ee }

\section{Construction of Ground states and quasiholes from parton structures}
We emphasize that the results of the preceding two sections only impose necessary conditions on the existence
of zero modes of the Hamiltonian (1) of the main text: A priori, the existence of a pattern composed of the units and according to the rules established in the foregoing does not guarantee the existence of a zero mode whose root state is described by this pattern. Together with a construction principle for such zero modes, however, the EPP governing root states has far reaching consequences. In particular, if for every allowed dominance pattern a zero mode can be constructed whose root state precisely corresponds to this pattern, it follows that the wave functions so constructed are a {\em complete} set of zero modes. This has been established by some of us earlier\cite{ortiz} and generalizes effortlessly to the present, multi-Landau-level context\cite{Chen}. We will apply this reasoning now to the case at hand. Consider thus wave functions of the form (2) of the main text, or
\begin{equation}\label{ZM2}
\psi=\prod_{i<j}(z_i-z_j) \,D_1\,D_2\,,
\end{equation}
where $D_1$ and $D_2$ can be taken to be Slater-determinants consisting only of the {\em first two} types of orbitals in \Eq{mono}. In this section, we again find it advantageous to work with the single particle basis \eqref{mono}, and omit all Gaussian factors. If now we take $D_1=D_2$ equal to the ``densest'' Slater-determinant $\cal D$, where for some $N$, all orbitals admissible orbitals $z^j$ ($j\geq 0$) and $\bar z z^{j+1}$ ($j\geq -1$) with $j\leq (N-1)/2$ are occupied then the resulting zero mode is dominated by the root state with the densest dominance pattern, i.e. $1_2001100110011\dotsc$; here the subscript 2 indicates that the leading particle resides in the second excited Landau level (as it must, having $j=-2$).  The pattern is ``densest'' in the sense that for a given $N$, no pattern of smaller total angular momentum is possible, nor any pattern whose largest occupied orbital has smaller (single particle) angular momentum. It immediately follows that the zero mode with $D_1=D_2={\cal D}$ {\em is} the densest zero mode as conjectured earlier in \cite{jain221}, for any odd $N$. For, any zero mode of the same $N$ but smaller total angular momentum or smaller highest occupied orbital would necessarily have a root states with the same properties, and this root state could then not satisfy the EPP.

This reasoning can be extended to show that the zero modes \eqref{ZM2} form a(n) (over-)complete set of zero modes. In algebraic terms, this proves the quite non-trivial theorem that the set of all polynomials in $z_i, \bar z_i$, with the requisite anti-symmetry, at most second order in any $\bar z_i$, {\em and} having at least third order zeros as $z_i\rightarrow z_j$, $\bar z_i\rightarrow \bar z_j$ is already linearly generated by the states of the form \eqref{ZM2}, i.e., Jastrow-factor times a product of two Slater determinants in $z_i$, $\bar z_i$, {\em each} at most linear in any $\bar z_i$. Clearly, this statement has useful generalizations to other parton states involving higher Landau levels and similarly constructed parent Hamiltonians, which we will leave for future work.

The detailed argument proceeds as follows. Below we construct for every dominance pattern $d$ allowed by the EPP a state $\psi_d$ of the form \eqref{ZM2} such that the root state of $\psi_d$ is precisely $\ket{d}$, i.e., the root state associated with the pattern $d$. The construction is such that $\braket{d'|\psi_d}$ may be non-zero for some $d'\neq d$, however, the matrix  $\braket{d'|\psi_d}$ will have a triangular structure with non-zero diagonal, and thus be invertible. This ensures the completeness of the $\psi_d$. For one, it trivially implies the linear independence of the $\psi_d$. What's more, to any zero mode $\ket{\phi}$ we may then construct a linear combination $\ket{\tilde\phi}$ of the $\psi_d$ such that $\braket{d|\tilde\phi}=\braket{d|\phi}$ for all root states $\ket{d}$ allowed by the EPP. This means that $\ket{\tilde\phi}-\ket{\phi}$ is a zero mode that's orthogonal to all permissible root states. This is only possible if $\ket{\phi}=\ket{\tilde\phi}$. Thus $\ket{\phi}$ is already a linear combination of the $\psi_d$.

We proceed with the construction of $\psi_d$. We introduce the short hand notation ${(\ell)_i}^n=z_i^{\ell+n}\bar z_i^{n}$ for the monomials \eqref{mono}, not including the normalization, which is inessential for present purposes.
We will use the notation
\begin{equation}
\{{(\ell_1)_1}^{n_1}\dotsc {(\ell_N)_N}^{n_N}\}
\end{equation}
for anti-symmetrized products of these monomials, where we will always insist that $\ell_i\leq \ell_{i+1}$. The $D_1$, $D_2$ in \eqref{ZM2} are of this form, with the additional constraint that $n_i\leq 1$. There is a simple rule describing ``dominance'' for a product of two Slater-determinants of this form, first stated for the lowest LL case \cite{regnault} ($n_i=0$), but easily generalized to $n_i\geq 0$ \cite{Chen}. This is that in the expansion of the product of  $\{{(\ell_1)_1}^{n_1}\dotsc {(\ell_N)_N}^{n_N}\}$ and $\{{(\ell'_1)_1}^{n'_1}\dotsc {(\ell'_N)_N}^{n'_N}\}$ into Slater determinants,
there is a non-expandable Slater determinant of the form $\{{(\ell_1+\ell'_1)_1}^{n_1+n'_1}\dotsc {(\ell_N+\ell'_N)_N}^{n_N+n'_N}\}$.
The key novel feature for multiple LLs is that while the rule $\ell_i\leq \ell_{i+1}$, $\ell'_i\leq \ell'_{i+1}$ fixes the angular momenta $\ell_i+\ell'_i$ of ``dominant'' (non-expandable) Slater determinants in the product, in the case of multiple degenerate $\ell_i$, the order of the associated $n_i$ is arbitrary. The dominance-rule can be applied to any such ordering, leading to all the different non-expandable Slater determinants in the product, all of which have the same angular momentum quantum numbers or occupied lattice positions, but differ in the LL-related indices $n_i+n'_i$.  This phenomenon precisely leads to the root state entanglement we know to be required, in general, of zero modes!
The rule can be straightforwardly generalized to products of three Slater determinants. Note that one may write the Jastrow-factor in \eqref{ZM2} as ${\cal J}=\{(0_1)_0 (1_2)_0\dots((N-1)_N)_0\}$, making this rule straightforwardly applicable to \Eq{ZM2}. Table \ref{ruletable} shows how any of the three building blocks of the EPP can be mapped onto units in $D_1$ and $D_2$ such that the root state of ${\cal J}D_1D_2$ will contain this building block at the right position. It is worth considering the $1_{\sigma_L}01\dotsc 1_{\sigma_R}$ block. The product rule described above when applied to ${\cal J}D_1 D_2$ as given in the table readily implies that the resulting orbital pattern at root level, without regard to LL-indices, is $101\dotsc 1$. One may now argue that the rule of Table \ref{ruletable} results in the AKLT-type MPS structure described in the proceeding sections in two slightly different ways.
One may check directly that the permissible permutations of the $n$, $n'$-indices described above reproduce the advertized MPS structure. Alternatively, it is sufficient to point out that, all other parts of $D_1$ and $D_2$ staying the same, the rule of Table \ref{ruletable} results in four linearly independent zero modes with the $101\dotsc 1$ orbital pattern at root level. By the necessary criteria of the proceeding two sections, the entanglement structure at root level {\em must} then be consistent with the four AKLT-MPS states (or linearly independent linear combinations thereof).

We have thus constructed a set of zero modes $\{\psi_d\}$ of the form \eqref{ZM2}, where for any dominance pattern $d$ conforming to the EPP, $\psi_d$ is dominated by the root state $\ket{d}$ associated to $d$. To establish the completness property of these zero modes, as explained above, we need only consider the matrix $\braket{d'|\psi_d}$. We follow the argument of \cite{Chen}. Diagonal elements are non-zero by construction.
Moreover, for $\braket{d'|\psi_d}$ to be non-zero for some $d'\neq d$, $d'$ must be obtainable from $d$ by the inward-squeezing processes defined in the main text. Such processes always strictly decrease the value of the ``moment''
\begin{equation}
M=\sum_j \sum_n j^2 c^\dagger_{n,j} c_{n,j}\,,
\end{equation}
of which all $\ket{d}$ are eigenstates.
Thus, if we order the $\ket{d}$ according to increasing $M$, the matrix  $\braket{d'|\psi_d}$ is upper triangular, hence invertible. This completes the proof of the one-to-one correspondence between zero modes and dominance patterns.

\begin{table*}[h]
\begin{ruledtabular}
\begin{tabular}{  c | c | c  }
  EPP building block & $D_1$ & $D_2$  \\
  \hline
  \hline
  $\dotsc 1_{s_z}\dotsc$  & $\{\dotsc {(\lfloor\frac{j-i+1}{2}\rfloor )_i}^{\lfloor\frac{s_z+1}{2}\rfloor}\dotsc\}$  & $\{\dotsc {(\lceil\frac{j-i+1}{2}\rceil )_i}^{\lceil\frac{s_z+1}{2}\rceil}\dotsc\}$   \\
    \hline
  $\dotsc 11\dotsc$  & $\{\dotsc (\lfloor\frac{j-i+1}{2}\rfloor )_i^{0}\,(\lfloor\frac{j-i+1}{2}\rfloor )_{i+1}^{1}\dotsc\}$  & $\{\dotsc (\lceil\frac{j-i+1}{2}\rceil )_i^{0}\,(\lceil\frac{j-i+1}{2}\rceil )_{i+1}^{1}\dotsc\}$   \\
    \hline
  $\dotsc 1_{\sigma_L}01\dotsc 01_{\sigma_R}\dotsc$  & $\{\dotsc (\lfloor\frac{j-i+1}{2}\rfloor )_i^{n_1}\,(\lfloor\frac{j-i+2}{2}\rfloor )_{i+1}^{n_2}\dotsc (\lfloor\frac{j-i+k}{2}\rfloor )_{i+k}^{n_k}\dotsc$\}  & $\{\dotsc (\lceil\frac{j-i+1}{2}\rceil )_i^{n_1'}\,(\lceil\frac{j-i+2}{2}\rceil )_{i+1}^{n_2'}\dotsc(\lceil\frac{j-i+k}{2}\rceil )_{i+1}^{n_k'}\dotsc\}$
\end{tabular}
\end{ruledtabular}
\caption{\label{ruletable}Rules for distributing the building blocks of the EPP over corresponding units in $D_1$ and $D_2$, \Eq{ZM2}. The leading particle in the EPP block is assumed to be the $i$th particle and occupying the orbital with angular momentum $j$. A ``free'' $1_{s_z}$-block leads to two singly occupied orbital in both $D_1$ and $D_2$ (see text for notation). A $11$-block leads to double occupied orbitals in both $D_1$ and $D_2$. In a $101$-block of $k$-particles, third row, there is no real freedom in choosing most of the $n_{1\dots k},\;  n_{1\dots k}'\in \{0,1\}$, as most orbitals will be doubly occupied, for both $D_1$ and $D_2$. However, for both $k$ even and odd, among $D_1$ and $D_2$ there will be exactly one singly occupied orbital at the left, and exactly one singly occupied orbital at the right. I.e., among the $n_1, n_1'$, exactly one is free, say $n_1$, and may be identified with $\sigma_L$ via $n_1=\sigma_L+\frac 12$. The analogous statement holds for $n_k$, $n_k'$ and $\sigma_R$.
Observing that adjacent EPP-blocks in the same pattern are padded from one another by double zeros, it is easy to see the corresponding units in the Slater-determinants do not overlap in orbital space, both for $D_1$ and for $D_2$, respectively.}
\end{table*}
\section{Braiding Statistics from EPP}
One can demonstrate \cite{SeidelYang08,TTpapers} that the dominance patterns as defined here agree with the thin cylinder limiting form of analytic trial wave functions, and are of course likewise expected to agree with thin torus limits, as demonstrated in many cases(e.g.\cite{seidel05,seidel2006abelian}). Moreover, it is generally found that the thin torus limit of zero modes of parent Hamiltonians (such as Eq. (1) of the main text) is adiabatically connected to zero modes of a ``thick'' (therefore, two-dimensional) torus. This adiabatic continuity can be exploited via the ``coherent state method'' to extract braiding statistics from microscopic rules governing dominance patterns \cite{flavinPRX}, here the EPP. We will present some key steps of this method as applied to the present case, and leave details, regarding statistics and general torus wave functions, for future publications.

As argued in the main text, the topological information ingrained in the EPP for the Jain-221 state is highly analogous to similar data for the $\nu=1/2$ Moore-Read state. Hence, the task is essentially to generalize earlier discussions \cite{seidel08, flavinPRX} for bosons at $\nu=1$ to fermions at $\nu=1/2$.
The heart of the method is a ``topological table'' as given by Table. \ref{toptable}.
This table illustrates how features of dominance patterns associated with states of few quasiholes (here: two) determine a coherent state Ansatz that is used to extract phases associated with two basic types of operations. Translations (T) describe transitions between different ``types'' or topological sectors under orbital (magnetic) translations. The rightmost column (F) describes exchange processes between quasiholes along topologically non-trivial paths.
In the table, patterns are shown without the spin-1/2 degrees of freedom, which we choose identical for all domain-walls (represented as $|$ for additional clarity) associated to quasi-holes. F operations translate the first domain $(|_1)$ wall to the second one $(|_2)$, while the latter will be translated to the position of first domain wall around one of the ``holes'' of the torus.
\begin{table}[H]
\begin{small}
 \begin{tabular}{c |c |c |c}
  \hline
 \hline
Sector & Domain walls & T & F\\
 \hline
 1 & 1010$\rvert_1$01100110$\rvert_2$01010 & 1,2 & $1,5+2\eta$ \\
 2 & 01010$\rvert_1$01100110$\rvert_2$0101 & $(-1)^{1+\eta},3$ & $1,6+2\eta$ \\
 3 & 101010$\rvert_1$01100110$\rvert_2$010 & 1,4 & $(-1)^{1+\eta},7-2\eta$ \\
 4 & 0101010$\rvert_1$01100110$\rvert_2$01 & $(-1)^{1+\eta},1$ & $(-1)^{1+\eta},8-2\eta$ \\
 5 & 110$\rvert_1$010101010$\rvert_2$01100 & 1,6 & $(-1)^{1+\eta},2$ \\
 6 & 0110$\rvert_1$010101010$\rvert_2$0110 & 1,7 & $1,3$ \\
 7 & 00110$\rvert_1$010101010$\rvert_2$011 & $(-1)^{1+\eta},8$ & $1,4$ \\
 8 & 100110$\rvert_1$010101010$\rvert_2$01 & $(-1)^{1+\eta},5$ & $1,1$ \\
 \hline
 \hline
\end{tabular}
\caption{{Topological Table}. $\eta$ is the particle number parity. Column T shows the phase and the new sector, respectively, one gets once T is applied on the given sector to the left. Column to is analogous for F. For illustrative purposes, we note that $T|2\rangle=-(-1)^\eta|3\rangle$ where $|2\rangle$ denotes a coherent state in sector 2.
 \label{toptable}}
 \end{small}
\end{table}

Having identified and labeled topological sectors for two quasi-holes as in the table, we may now be interested in the braid matrix for the adiabatic exchange of two quasi-holes (these must be thought of as localized in space via the coherent stat Ansatz, see \cite{flavinPRX} for details). Locality imposes stringent constraints on what matrix element may in principle be non-zero. Generally, only those matrix elements can be non-zero whose associated patterns in Table \ref{toptable} differ only {\em in between} the domain walls, but not to the left or right of the domain walls \cite{flavinPRX}.
Moreover, taking into account translational symmetry  this dictates the following general structure of the braid matrix:
 \begin{eqnarray}
 \Gamma=
 \begin{pmatrix}
 a&0&b&0&0&0&0&0\\
 0&a&0&b&0&0&0&0\\
 b' &0&a&0&0&0&0&0\\
 0& b' &0&a&0&0&0&0\\
 0&0&0&0&c&0&0&0\\
 0&0&0&0&0&c&0&0\\
 0&0&0&0&0&0&c&0\\
 0&0&0&0&0&0&0&c\\
\end{pmatrix}
 \end{eqnarray}
One may piece together the information of Table \ref{toptable} with the above structure and additional requirements from S-duality \cite{seidel10} on the torus, following the method of \cite{flavinPRX}. This fixes all entries down to a set of eight possible solutions, all related by Abelian phases and complex conjugation. In particular, one finds $b=\pm i a$, $b'=-b^\ast$, which is essentially responsible for a description in terms of Majorana fermions as mentioned in the main text. The operation of braiding on patterns with more than two quasi-holes is generated by applying the rules given for two quasi-holes to any pair of neighboring domain-walls in the associated patterns. Details will be given in a future publication.
\end{document}